\documentclass[a4paper,onecolumn,3p,preprint]{elsarticle}

\usepackage{lineno,hyperref}
\modulolinenumbers[5]

\journal{Physica E}

\begin{document}

\title{Tuning the Fano factor of graphene via Fermi velocity modulation}

\author[UFRPE]{Jonas R. F. Lima}
\ead{jonas.lima@ufrpe.br}

\author[UFRPE]{Anderson L. R. Barbosa}
\ead{anderson.barbosa@ufrpe.br}

\author[UFRN]{C. G. Bezerra}

\author[UFRN]{Luiz Felipe C. Pereira}
\ead{pereira@fisica.ufrn.br}

\address[UFRPE]{Departamento de F\'{\i}sica, Universidade Federal Rural de Pernambuco, 52171-900, Recife, PE, Brazil}
\address[UFRN]{Departamento d e F\'isica, Universidade Federal do Rio Grande do Norte, 59078-970, Natal, RN, Brazil}

\cortext[cor1]{Corresponding authors}

\date{\today}

\begin{abstract}
In this work we investigate the influence of a Fermi velocity modulation on the Fano factor of periodic and quasi-periodic graphene superlattices. We consider the continuum model and use the transfer matrix method to solve the Dirac-like equation for graphene where the electrostatic potential, energy gap and Fermi velocity are piecewise constant functions of the position $x$. { We found that in the presence of an energy gap, it is possible to tune the energy of the Fano factor peak and consequently the location of the Dirac point, by a modulations in the Fermi velocity}. Hence, the peak of the Fano factor can be used experimentally to identify the Dirac point. { We show that} for higher values of the Fermi velocity the Fano factor goes below 1/3 in the Dirac point. { Furthermore, we show that in periodic superlattices  the location of Fano factor peaks is symmetry when the Fermi velocity $v_A$ and $v_B$ is exchanged, however by introducing quasi-periodicity the symmetry is lost}. The Fano factor usually holds a universal value for a specific transport regime, which reveals that the possibility of controlling it in graphene is a notable result.
\end{abstract}

\maketitle


\section{Introduction}

At low temperatures, the electrical current through nanostructure presents time dependent fluctuations due to the discreteness of the electrical charge \cite{Blanter20001,BEENAKKER20161,DEJONG1996219}. These fluctuations are known as shot noise, and give rise to the Fano factor \cite{PT}, which is defined as the ratio between the shot noise power and mean electrical current. The Fano factor holds universal values to each regime of electronic transport as in the case of tunnel junctions which gets a Poissonian noise { ($F=1$)}, diffusive wires which is  { ($F=1/3$)} \cite{PhysRevB.46.1889,NAGAEV1992103} and chaotic quantum dots which is  { ($F=1/4$)} \cite{Fujii2010}.

With the advent of graphene electronics \cite{GOPAR201623,Morozov,Stormer,refId0}, it was shown that the conductivity of a graphene strip gets a minimum value when the gate voltage vanishes, which means that the shot noise power goes to a maximum value. With this in mind, the author of Ref. \cite{PhysRevLett.96.246802} showed theoretically that the Fano factor is 1/3 for a graphene strip, which coincides with the value for diffusive wires. This result is a consequence of the non-classical dynamics in the graphene strip introduced by Dirac equation and it has been measured in Refs. \cite{PhysRevLett.100.196802,PhysRevLett.100.156801}. { Moreover}, the Fano factor has been studied in { carbon nanotubes \cite{PhysRevLett.99.036802,PhysRevLett.99.156803,PhysRevLett.99.156804,LuisTorres}}, graphene nanoribbons \cite{GOPAR201623} and superlattices, such as periodic \cite{doi:10.1063/1.3599447,doi:10.1063/1.3525270,Azarova2014,Sattari2015,Razzaghi2015} and quasi-periodic following a Double-periodic sequence \cite{0022-3727-46-1-015306}, the Fibonnaci sequence \cite{doi:10.1063/1.3658394} and the the Thue-Morse sequence \cite{doi:10.1063/1.4826643,doi:10.1063/1.4729133,doi:10.1063/1.4772209,doi:10.1063/1.4788676}. Furthermore, { Ref. \cite{PhysRevB.91.115135} proposed that the Fano factor of a graphene strip can get a correction if the Dirac cone holds a tilt  and Refs.  \cite{LuisTorres,PhysRevB.81.115435} showed that with time-dependent fields it is possible to minimize the electronic current fluctuations, as well as exploring scaling properties of shot-noise with the length and width of the graphene device. Therefore, Refs.\cite{LuisTorres,PhysRevB.91.115135,PhysRevB.81.115435} show that it is possible to control the Fano factor via experimental parameters in the Dirac equation.}

In recent years, the study of a Fermi velocity modulation on the electronic properties of graphene has attracted a great deal of interest, since the Fermi velocity of graphene can be engineered, for instance, by the presence of a substrate or doping \cite{Hwang,Attaccalite}. A position-dependent Fermi velocity can be obtained by placing metallic planes close to the graphene lattice, which will change the electron concentration in different regions\cite{Polini,Yuan}. It has been shown, for instance, that a Fermi velocity modulation can be used to control the energy gap of graphene \cite{Lima2}, to induce an indirect energy gap in monolayer \cite{Lima1} and bilayer graphene \cite{hosein}, to engineer the electronic structure of graphene \cite{Lima3,Lima4} and to create a waveguide for electrons in graphene \cite{Polini,Yuan}. Most recently, it was shown that it is possible to tune the transmitivity of electrons in graphene from 0 to 1 with the Fermi velocity, which can be used to turn on/off the electronic transport in graphene \cite{jrfl2016}.

Motivated by these studies, in this work we investigate the influence of a Fermi velocity modulation on the Fano factor of a graphene superlattice. We consider the continuum model and use the transfer matrix method to solve the Dirac-like equation for graphene where the electrostatic potential, energy gap and Fermi velocity are piecewise constant functions of the position $x$. We found that, without an energy gap, the Fano factor in the location of the Dirac cone equals 1/3 for values of the Fermi velocity $v_F$ that were already obtained experimentally. Only for much higher values of $v_F$ the Fano factor goes below 1/3. However, in the presence of an energy gap, it is possible to tune the value and the location energy of the peak of the Fano factor in the location of the Dirac point by a Fermi velocity modulation. We also investigated the influence of the periodicity of the graphene superlattice in the Fano factor.

\section{Methods}

The low-energy electronic states of single layer graphene can be described by a Dirac-like Hamiltonian given by $H=-i\hbar v_F \left(\sigma_x \partial_x +\sigma_y \partial_y\right)$, where $\sigma_i$ are the Pauli matrices acting on the \textit{pseudospin} related to the two graphene sublattices, and $v_F$ is the Fermi velocity. We consider that the superlattice has two regions, A and B, and that the electrostatic potential, energy gap and Fermi velocity in region A (B) are given by $V_A (V_B)$, $\Delta_A (\Delta_B)$ and $v_A (v_B)$, respectively. Thus, the effective Dirac Hamiltonian for a graphene superlattice with position-dependent Fermi velocity and energy gap is given by \cite{peres} $H=-i\hbar \left(\sqrt{v_F(x)}\sigma_x \partial_x  \sqrt{v_F(x)} +v_F(x)\sigma_y \partial_y\right) +V(x)\hat{1}+\Delta(x)\sigma_z$, 
where $V(x)$ is the electrostatic potential and $\Delta(x)$ is half the energy gap. The Dirac equation is given by $H\psi(x,y)=E\psi(x,y)$ and due to the translation invariance in the $y$ direction, we can write the wave function as $\psi(x,y)=e^{-ik_yy}\psi(x)$.

\begin{figure*}[htb]
\centering
\includegraphics[width=0.9\textwidth]{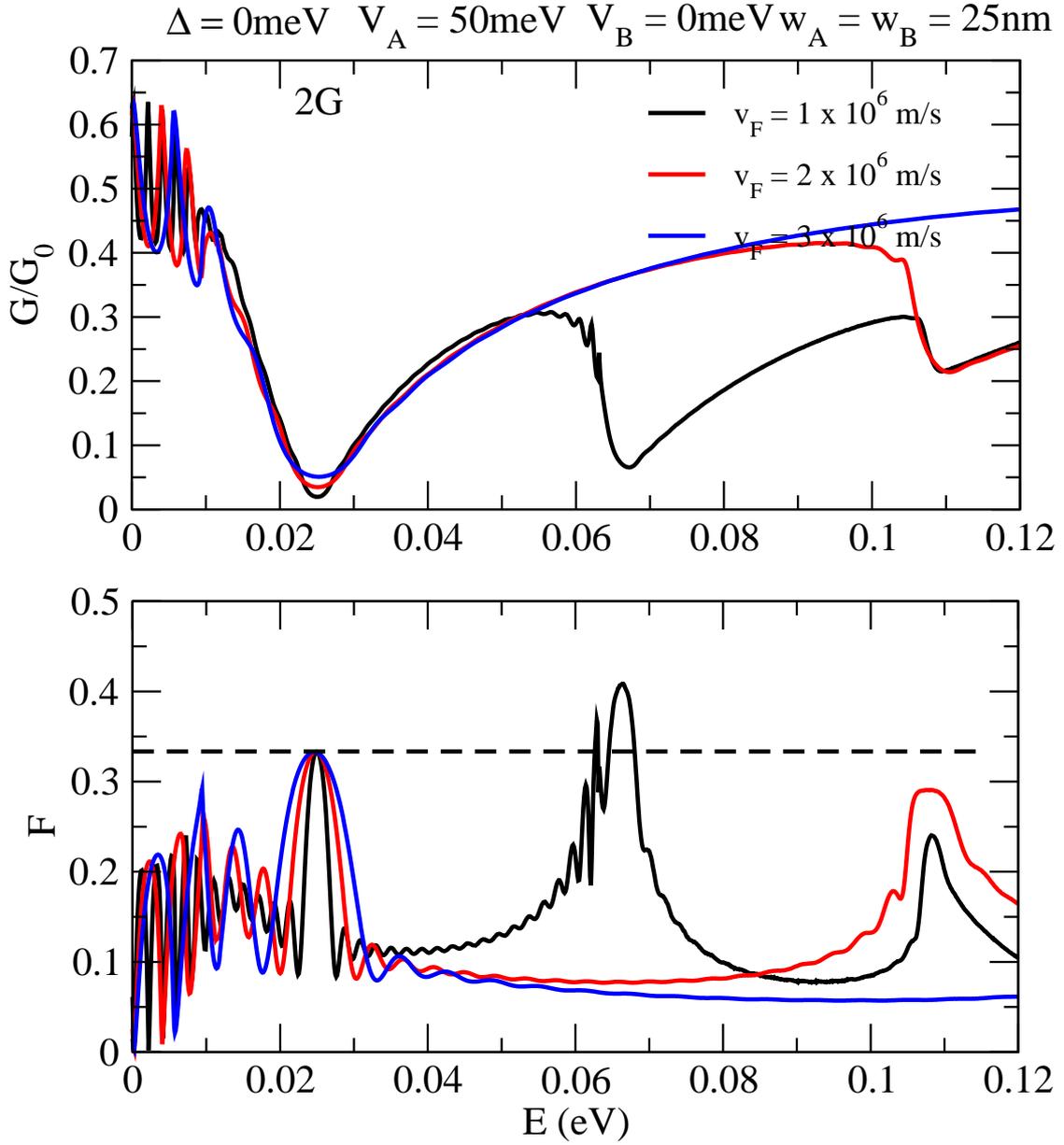}	
\caption{Conductance and Fano factor as a function of the energy for a constant Fermi velocity with $\Delta=0$. The horizontal dashed line represent $F=1/3$. The values of the parameters of the system are written in the figure. The location of the Dirac point is $E=25$ meV.}
\label{2G}
\end{figure*}

Using the transfer matrix method, we obtain that the transfer matrix connecting the wave function $\psi(x)$ at $x$ and $x+\Delta x$ in the $j$th barrier is given by \cite{Yao10}
\begin{equation}
M_j(\Delta x, E, k_y)=\left(
\begin{array}{cc}
\frac{\cos(q_j\Delta x - \theta_j)}{\cos\theta_j} & i\frac{\sin(q_j\Delta x)}{p_j \cos\theta_j} \\
i\frac{p_j \sin(q_j\Delta x)}{\cos\theta_j} & \frac{\cos(q_j\Delta x + \theta_j)}{\cos\theta_j}
\end{array} \right) \; ,
\end{equation}
where $p_j = \ell_j/k_j$ with $k_j = [(E-V_j)^2-\Delta_j^2]^{1/2}/(\hbar v_j)$ and $\ell_j = [(E-V_j)-\Delta_j]/(\hbar v_j)$. $q_j$ is the $x$ component of the wave vector and is given by $q_j = \sqrt{k_j^2-k_y^2}$ for $k_j^2>k_y^2$, otherwise $q_j = i\sqrt{k_y^2-k_j^2}$. $\theta_j$ is the angle between the $x$ component of the wave vector, $q_j$, and the wave vector, $k_j$, and is given by $\theta_j = \arcsin (k_y/k_j)$ when $|E-V_j| > \Delta_j$, otherwise $\theta_j = \arccos (q_j/k_j)$. The transmission coefficient can be written as
\begin{equation}
t(E, k_y) = \frac{2\cos \theta_0}{(x_{22}e^{-i\theta_0}+x_{11}e^{i\theta_e})-x_{12}e^{i(\theta_e-\theta_0)}-x_{21}} ,
\label{tc}
\end{equation}
where $x_{ij}$ are the elements of the total transfer matrix given by $X =  \prod_{j=1}^{N} M_j(w_j, E, k_y)$, for a superlattice with N regions. In possession of the transmission coefficient, one can calculate the total conductance of the system at zero temperature via the Landauer-B\"uttiker formula, which gives 
\begin{equation} 
G = G_0 \int_{-\pi/2}^{\pi/2} T \cos \theta_0 d\theta_0, 
\end{equation}
where $G_0 = 2e^2 E L_y /(\pi \hbar)$ and $T = |t|^2$ is the transmitivity. $L_y$ is the sample size in the $y$ direction. Finally, the Fano factor is calculated from 
\begin{equation}
F = \frac{\int_{-\pi/2}^{\pi/2} T (1-T) \cos \theta_0 d\theta_0}{\int_{-\pi/2}^{\pi/2} T \cos \theta_0 d\theta_0}.
\end{equation}

In order to analyze if the periodicity of the graphene superlattice affects how the Fermi velocity modulation influences the Fano factor, we will consider superlattices with a unit cell following a Fibonacci sequence for different generations. The Fibonacci sequence is obtained by the recurrence relation $S_{j+1} = \{ S_j, S_{j-1} \}$, with $S_0 = \{ B \}$ and $S_1 = \{ A \}$, where $j$ is the Fibonacci generation. Thus we have $S_2 = \{ AB \}$, $S_3 = \{ ABA \}$ and $S_4 = \{ ABAAB \}$. In what follows, we will consider $w_A = w_B = 25$ nm, $\Delta_A = \Delta_B = \Delta$ and $V_B = 0$. 

\begin{figure}[hpt]
\centering
\includegraphics[width=0.9\linewidth]{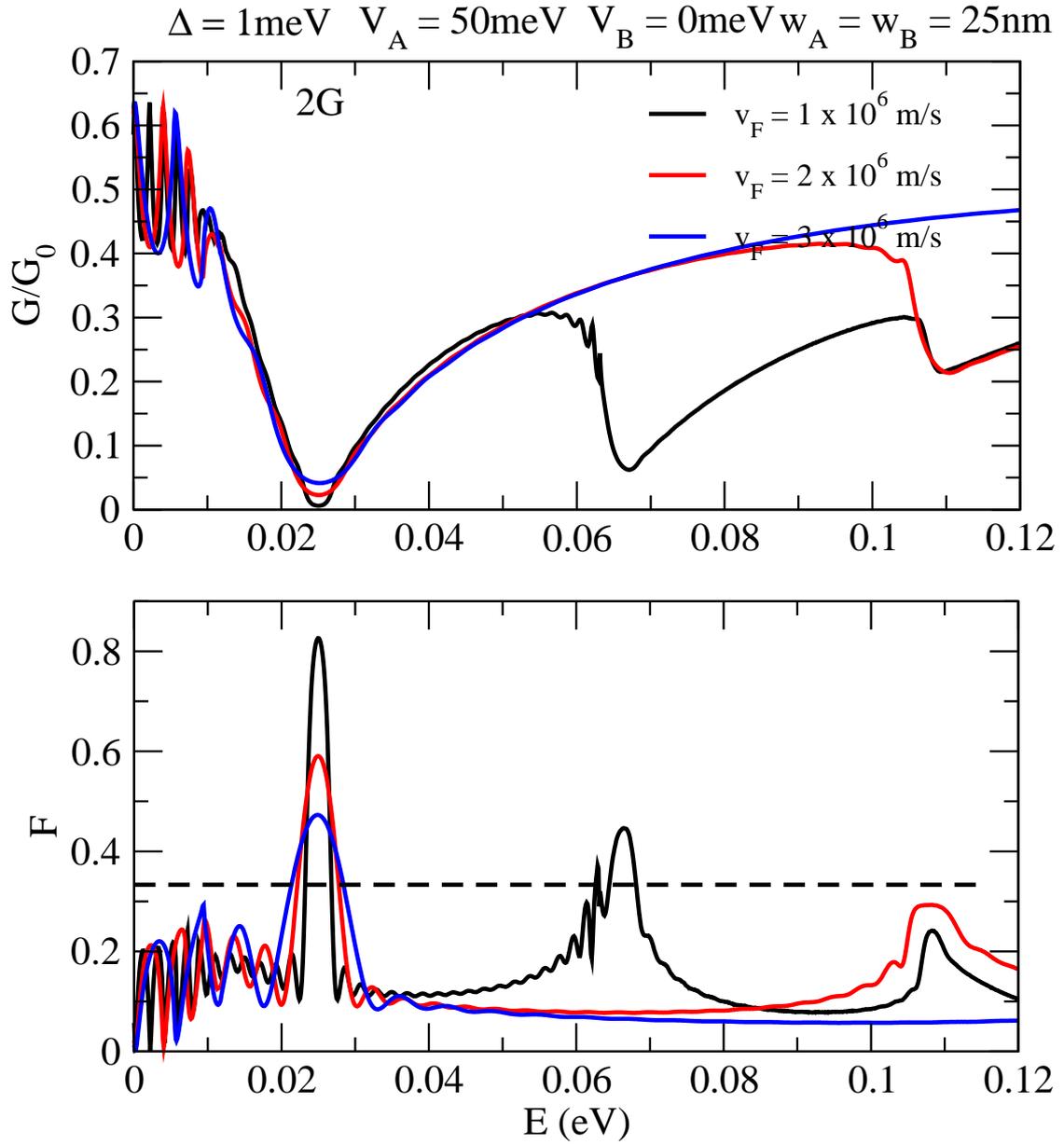}	
\caption{Conductance and Fano factor in terms of the energy of the system for a constant Fermi velocity with $\Delta=1$ meV. The horizontal dashed line represent $F=1/3$. The values of the parameters of the system are written in the figure.}
\label{2Gd}
\end{figure}

\section{Results}

\subsection{Second Fibonacci generation}

In Fig. \ref{2G} we consider the second generation (2G) of the Fibonacci sequence, which is the periodic case. It presents the conductance and the Fano factor for three different values of $v_F$, considering the Fermi velocity constant in the whole superlattice and $\Delta = 0$. As we expect, one can see that a minimum of conductance coincides with a maximum of the Fano factor. One can also note that it is possible to change the amplitude and the location of some peaks of the Fano factor by changing the Fermi velocity.

The region of major importance is in the vicinity of the Dirac point. The location of the Dirac point is obtained from the relation $k_A + k_B = 0$. With a constant Fermi velocity, the exact location is  $E = V_A/2$ \cite{jonasjap}. It can be seen that the minimum value of the conductance occurs at the Dirac point, and the Fano factor is equal to $1/3$, independent of the value of the Fermi velocity, which is a robust feature in graphene.

\begin{figure}[htb]
\centering
\includegraphics[width=0.9\linewidth]{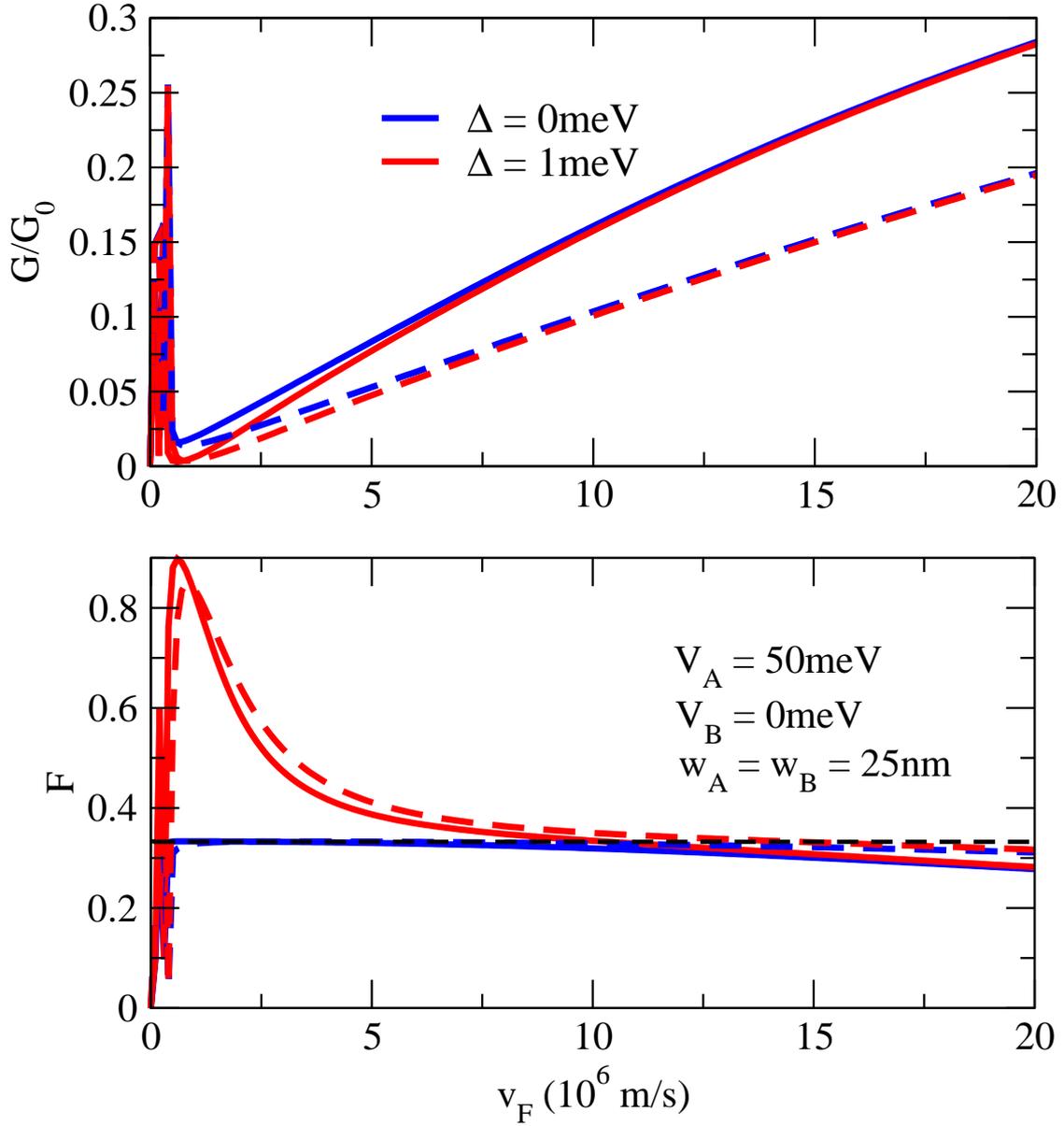}	
\caption{Conductance and Fano factor in terms of the Fermi velocity ($v_F$) for a constant incidence energy $E=25$ meV. The solid lines represent the second generation case, while the dashed lines represent the third generation case. The horizontal dashed line in the bottom panel represents $F=1/3$. }
\label{2GFV}
\end{figure}

In Fig. \ref{2Gd} we consider $\Delta = 1$ meV. In this case, the conductance and the Fano factor are almost the same as in the case with $\Delta = 0$. The only difference is at the Dirac point, where the Fano factor is no longer equal to $1/3$. This behavior can be understood since the conductance tends to decrease in the presence of an energy gap, which corresponds to an increase in the Fano factor. One interesting feature here is that, even though the energy gap increases the maximum value of the Fano factor, it is possible to reduce its amplitude by controlling the Fermi velocity. So, it is possible to compensate the presence of an energy gap with an increase in Fermi velocity.

The behavior observed in Fig. \ref{2Gd} raises one question: Is it possible to recover the value of $1/3$ for the Fano factor changing the Fermi velocity? The answer for this question is yes, as we show in Fig. \ref{2GFV}, where we present the conductance and the Fano factor at the Dirac point as a function of the Fermi velocity for $\Delta=0$ (black line) and $\Delta=1$ meV (red line). For $\Delta=1$ meV, the Fano factor equals $1/3$ when the Fermi velocity is around $10^7$ m/s, which is one order of magnitude greater than its usual value. For higher values of $v_F$, the Fano factor decreases and takes  values below $1/3$. This also happens when $v_F$ is higher than $5\times10^6$ m/s for the case with $\Delta=0$. 

\begin{figure}[hpt]
\centering
\includegraphics[width=0.9\linewidth]{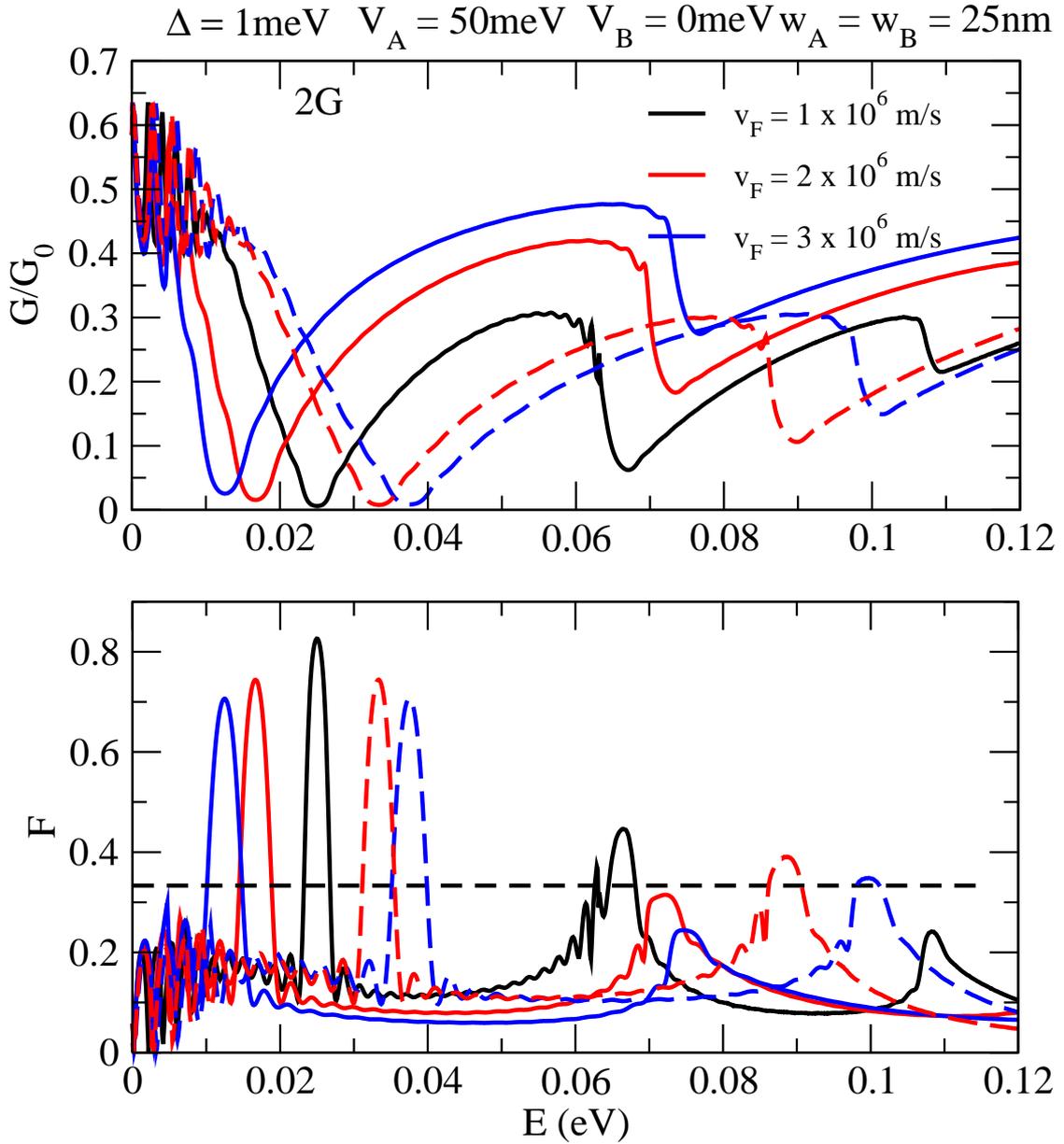}	
\caption{Conductance and Fano factor as a function of the energy with a modulation of the Fermi velocity. The continuum (dashed) lines represent the case with a fixed Fermi velocity of $1 \times 10^6$ m/s in the region $B$ ($A$), with the different colors representing different values of the Fermi velocity in the region $A$ ($B$). The horizontal dashed line represent $F=1/3$. The values of the other parameters of the system are written in the figure.}
\label{2Gdv}
\end{figure}

We consider a modulation in the Fermi velocity in Fig. \ref{2Gdv}, where we keep the Fermi velocity in region $B$ ($A$) equal to $1 \times 10^6$ m/s and consider three different values for the Fermi velocity in region $A$ ($B$), which is shown by the solid (dashed) lines. We notice that in this case, in contrast to the constant Fermi velocity case, the modulation of the Fermi velocity changes the amplitude of the Fano factor as well as their location in energy.

Looking at the vicinity of the Dirac point, our results become clearer. For $v_A\neq v_B$, the location of the Dirac point is given by 
\begin{equation}
E= \frac{V_Av_B^2-\sqrt{v_A^2 v_B^2(V_A^2-2\Delta^2)+\Delta^2(v_A^4+v_B^4)}}{v_B^2-v_A^2},
\end{equation}
which means that one can control the location of the Dirac point by changing the Fermi velocity and, consequently, control the location of the Fano factor peaks.

\begin{figure*}[htb]
\centering
\includegraphics[width=0.9\linewidth]{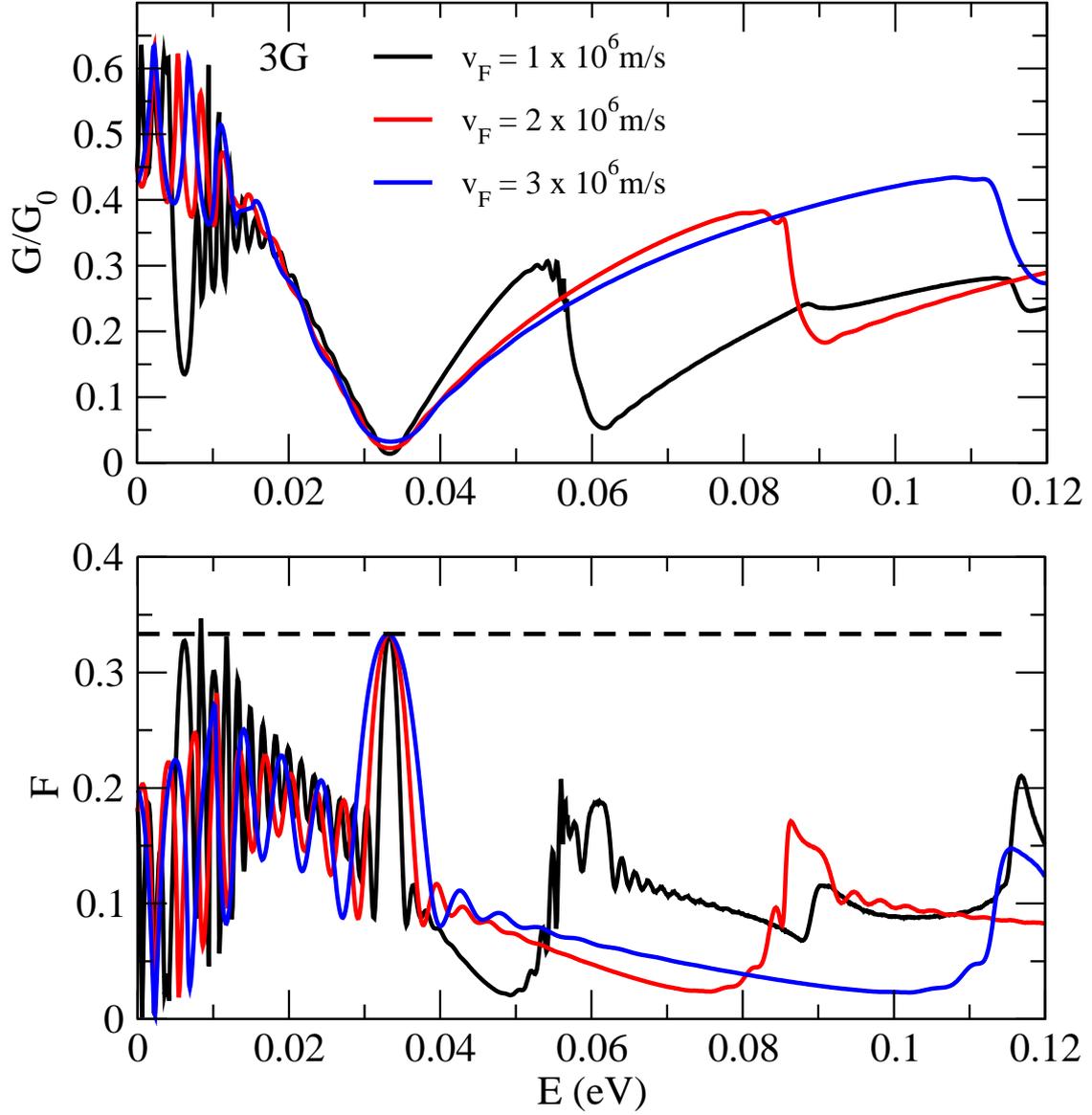}	
\caption{Conductance and Fano factor for the third generation of the Fibonacci sequence with a constant Fermi velocity and $\Delta=0$. The horizontal dashed line represent $F=1/3$. The values of the parameters of the system are the same as in Fig. \ref{2G}.}
\label{3G}
\end{figure*}

Thus, we can conclude that, without an energy gap, the Fermi velocity cannot change the robust features of the conductance and the respective Fano factor in the vicinity of the Dirac point. However, the introduction of an energy gap destroys the Dirac point, breaking this robust characteristic, which makes it possible to control the amplitude and the location of the peaks of the Fano factor by controlling the Fermi velocity.

\subsection{Third Fibonacci generation}

In Fig. \ref{3G} we consider the third generation (3G) of the Fibonacci sequence, which is a quasi-periodic sequence. We consider a constant Fermi velocity and $\Delta = 0$. As in the 2G case, at the Dirac point the Fano factor equals to $1/3$, independent of the Fermi velocity, and the conductance has its minimal value. For the 3G case, the location of the Dirac point with $v_A = v_B$ is given by 
\begin{equation}
E=\frac{4}{3}V_A - \frac{1}{3}\sqrt{4V_A^2+9\Delta^2}. 
\end{equation}
In contrast to the periodic case, the location of the Dirac point now depends on the energy gap $\Delta$.

\begin{figure}[hpt]
\centering
\includegraphics[width=0.9\linewidth]{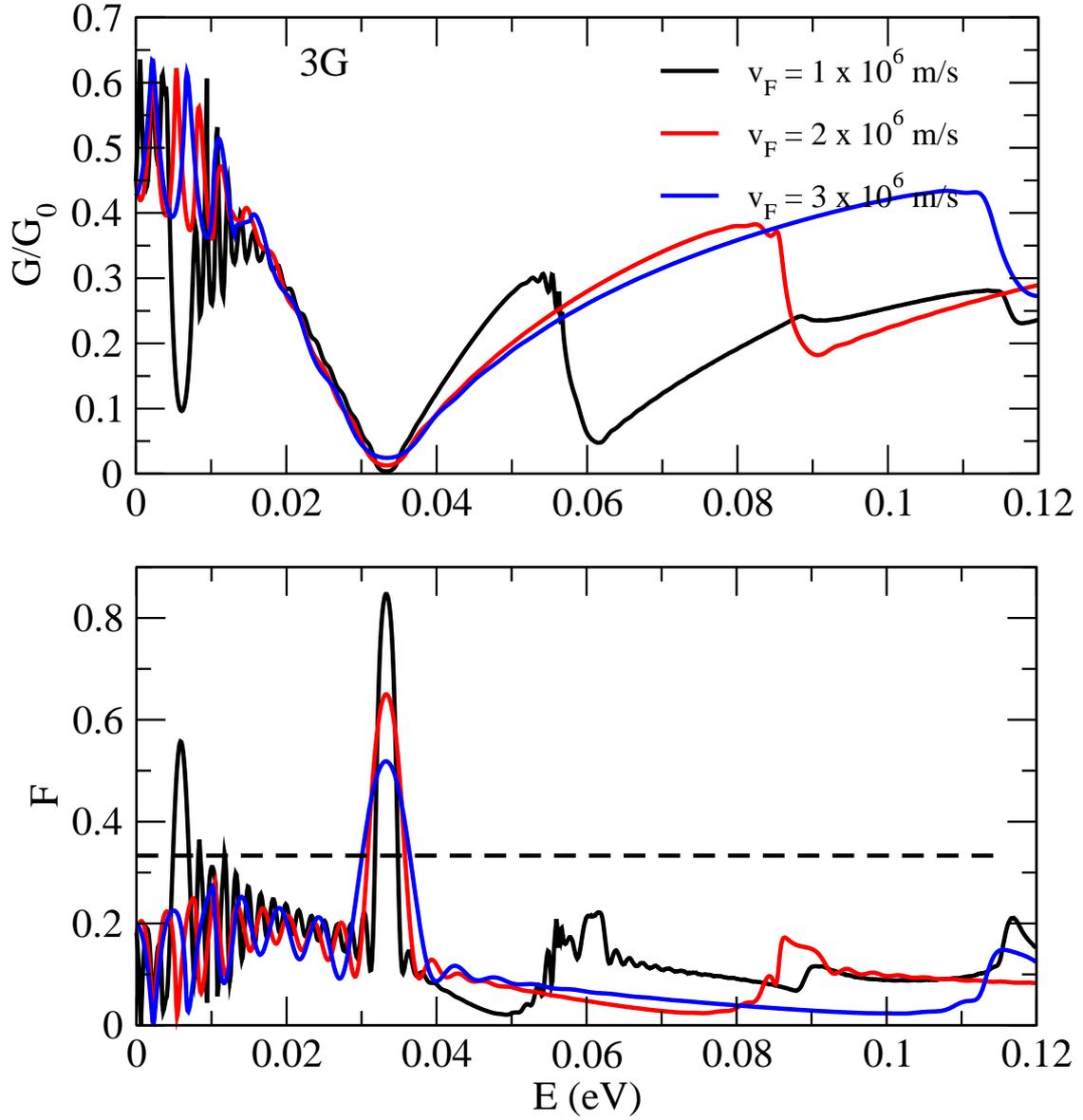}	
\caption{Conductance and Fano factor in terms of the energy of the system for the third generation of Fibonacci with a constant Fermi velocity and $\Delta=1$ meV. The horizontal dashed line represent $F=1/3$. The values of the parameters of the system are the same as in Fig. \ref{2Gd}.}
\label{3Gd}
\end{figure}

In Fig. \ref{3Gd} we consider $\Delta = 1$ meV. Since we are considering a small energy gap, the location of the Dirac point is only slightly smaller than in the case with $\Delta = 0$. Similar to the periodic case, the Fano factor peak in the Dirac point is very sensitive to the energy gap and the increase of the Fermi velocity reduces the effect of the energy gap, since the Fano factor peak gets closer to 1/3. The value of the total conductance and Fano factor at the Dirac point for higher values of $v_F$ can be seen by the dashed lines in Fig. \ref{2GFV}.

\begin{figure}[hpt]
\centering
\includegraphics[width=0.9\linewidth]{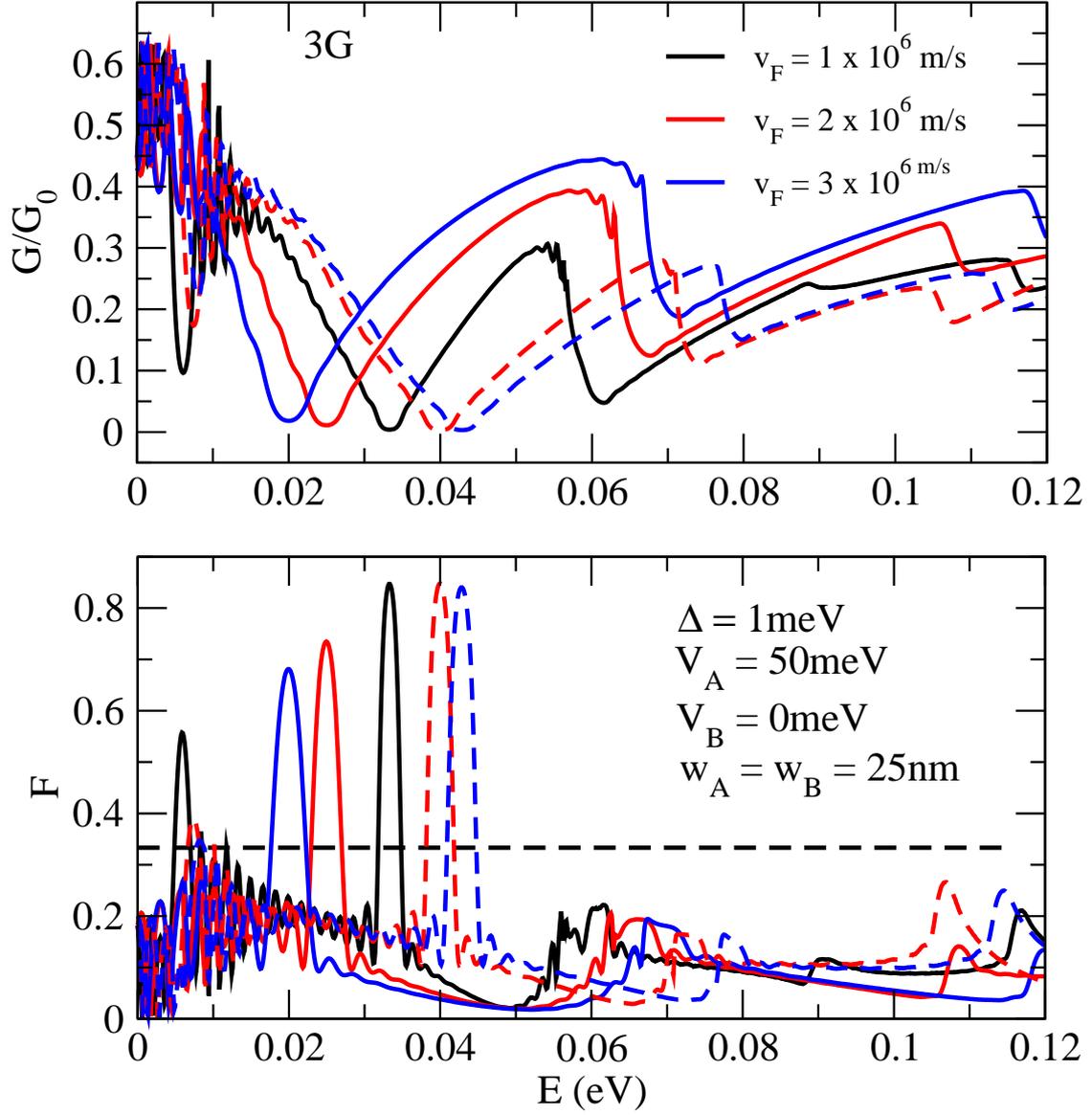}	
\caption{Conductance and Fano factor as a function of the energy for the third generation of Fibonacci with a modulation of the Fermi velocity. The continuum (dashed) lines represent the case with a fixed Fermi velocity of $1 \times 10^6$ m/s in the region $B$ ($A$), with the different colors representing different values of the Fermi velocity in the region $A$ ($B$). The horizontal dashed line represent $F=1/3$. The values of the other parameters of the system are the same as in Fig. \ref{2Gdv}.}
\label{3Gdv}
\end{figure}

We consider a position-dependent Fermi velocity in Fig. \ref{3Gdv}, where, as in the periodic case, we keep the Fermi velocity in region $B$ ($A$) equal to $1 \times 10^6$ m/s and consider three different values for the Fermi velocity in region $A$ ($B$), which is shown by the solid (dashed) lines. As in the periodic case, a modulation of the Fermi velocity changes the amplitude of the Fano factor peak at the Dirac point and controls its location in energy. However, the symmetry related to the peak for the case $v_A=v_B$, when we change $v_A$ or $v_B$, is lost. It is a consequence of the fact that in the unit cell that follows the third generation of the Fibonacci sequence, there are two regions $A$ and only one region $B$. So, the location of the Dirac point and the amplitude of the peak of the Fano factor is more sensitive to a change in the Fermi velocity in region $A$. With $v_A\neq v_B$, the location of the Dirac point is given by
\begin{equation}
E= \frac{4V_Av_B^2-\sqrt{4v_A^2 v_B^2(V_A^2-2\Delta^2)+\Delta^2(v_A^4+16v_B^4)}}{4v_B^2-v_A^2}.
\end{equation}

It is possible to find a general expression for the energy location of the Dirac point for any Fibonacci sequence when $v_A\neq v_B$. It is given by the relation $N_A k_A+N_Bk_B=0$, where $N_A$ and $N_B$ are the number of regions $A$ and regions $B$ in the Fibonacci sequence, respectively. With this relation, we obtain 
\begin{equation}
E= \frac{N_B^2 V_Av_B^2-\sqrt{N_A^2 N_B^2 v_A^2 v_B^2(V_A^2-2\Delta^2)+\Delta^2(N_A^4 v_A^4+N_B^4 v_B^4)}}{N_B^2 v_B^2-N_A^2 v_A^2}.
\label{gl}
\end{equation}
It is important to remember that in this expression for the location of the Dirac point we consider $w_A=w_B$ and $\Delta_A=\Delta_B=\Delta$. Without this assumptions, the equation would be different. It is possible to verify that if one replaces the values of $N_A$ and $N_B$ in Eq. \ref{gl} related to the two cases considered until now, the previous energy expressions are recovered.

\subsection{Fourth Fibonacci generation}

\begin{figure}[htb]
\centering
\includegraphics[width=0.9\linewidth]{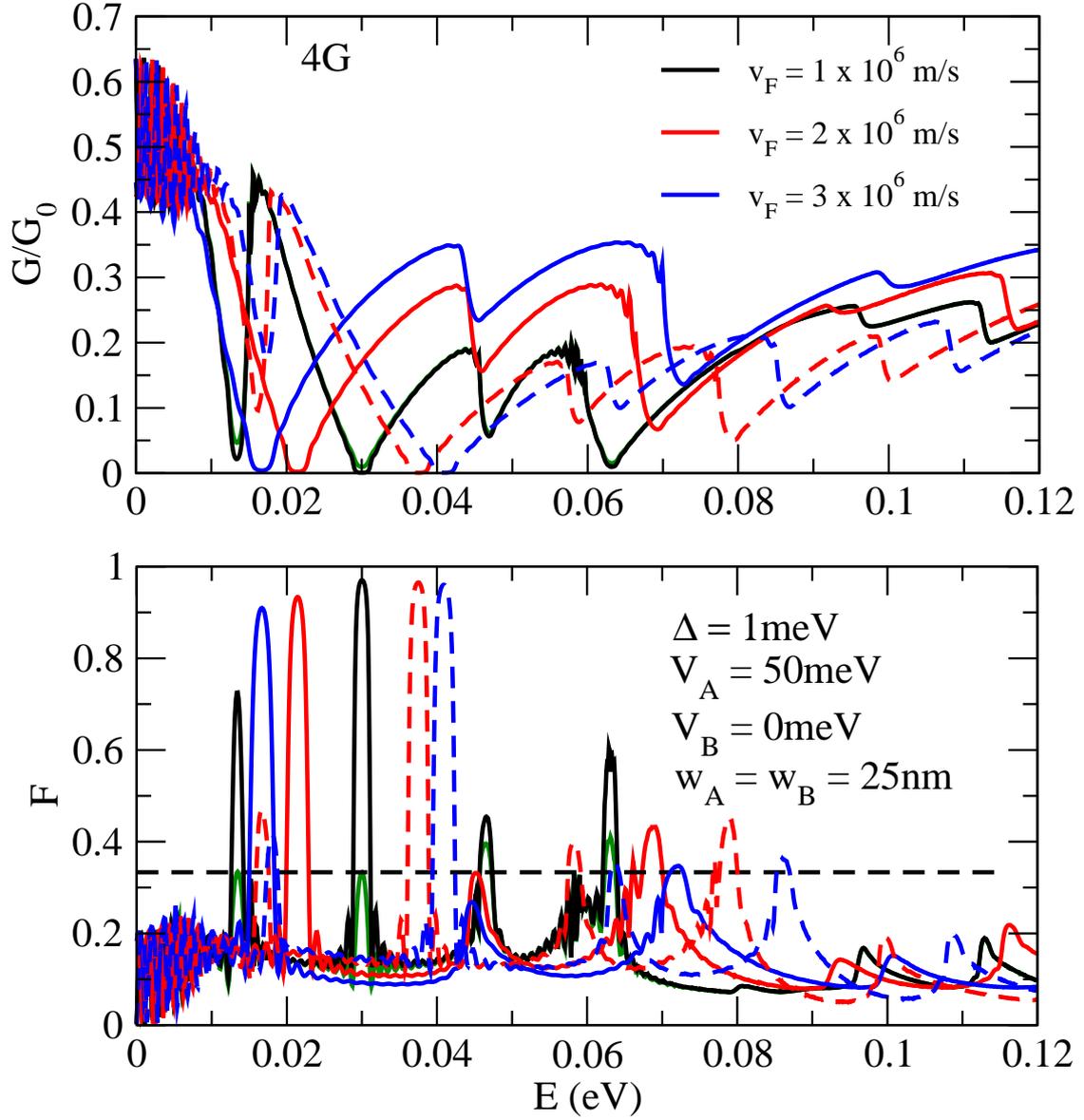}	
\caption{Conductance and Fano factor as a function of the energy with a modulation of the Fermi velocity. The green line represent the case with $\Delta=0meV$. The continuum (dashed) lines represent the case with a fixed Fermi velocity of $1 \times 10^6$ m/s in the region $B$ ($A$), with the different colors representing different values of the Fermi velocity in the region $A$ ($B$). The horizontal dashed line represent $F=1/3$. The values of the other parameters of the system are written in the figure.}
\label{4G}
\end{figure}

In order to verify how the different generations bring a new behavior to the system, we also consider the fourth generation of the Fibonacci sequence. In Fig. \ref{4G} we consider the 4G case with the same parameters as in Fig. \ref{2Gd} and, additionally, the green line represent the case with $v_A=v_B=1\times 10^6$ m/s and $\Delta=0$. The conductance and Fano factor for the fourth generation present the same behavior as in the third generation. This is caused by the different number of A and B regions in the 3G case. Since the fourth generation is given by ABAAB, the location of the Dirac point for this case is obtained replacing $N_A=3$ and $N_B=2$ in Eq. (\ref{gl}).

The ratio between the number of B regions and A regions in the unit cell of the graphene superlattice equals $1$ for the periodic case, $1/2$ for the 3G case and $2/3$ for the 4G case. As the Fibonacci sequence increases, this ratio gets closer and closer to $[1+\sqrt{5}]/2$, and the symmetry of the Fano factor peaks related to the case $v_A=v_B$ when we change $v_A$ or $v_B$, which was observed in the periodic case, can be recovered. 

\section{Conclusion}

In summary, we investigate the influence of a Fermi velocity modulation in the conductance and Fano factor of graphene superlattices. We verified that, without an energy gap, it is not possible to change the value of $1/3$ of the Fano factor at the Dirac cone of graphene by a realistic Fermi velocity modulation. The Fano factor becomes below $1/3$ only for very high Fermi velocity values, which were never obtained experimentally. However, in the presence of an energy gap, we showed that the Fermi velocity modulation can tune the value of the Fano factor in graphene and also control the location of its peaks, revealing that the Fano factor is very sensitive to the Fermi velocity. We also analyzed if the periodicity of the superlattice affects how the Fermi velocity modulation influences the Fano factor and we obtained that without periodicity, some symmetries are lost. These results turn possible to use the Fano factor as an indirect measurement of, for instance, the Dirac point location and energy gap of graphene. We consider a Fibonacci sequence, but the influence of the Fermi velocity modulation in the Fano factor obtained here should be the same for graphene superlattices following others non-periodic sequences, such as Double-periodic and Thue-Morse. The Fano factor usually holds a universal value for a specific transport regime, which reveals that the possibility of controlling it in graphene is a notable result. Since Fermi velocity engineering in graphene is a topic with important research advances in the last years, the results obtained here are useful for future applications of graphene in electronic devices. { As a highlight perspective, we can introduce electron-electron interactions in our model and perform a direct connection with the experiments of \cite{PhysRevLett.99.036802,PhysRevLett.99.156803,PhysRevLett.99.156804} and theories of \cite{LuisTorres,PhysRevB.81.115435,PhysRevB.84.115312}. }

\section{Acknowledgements}
The authors thank Tommaso Macr\`i for a critical reading of the manuscript. JRFL, ALRB, CGB were supported by CNPq. CGB and LFCP acknowledge financial support from Brazilian government agency CAPES for project ``Physical properties of nanostructured materials'' (Grant 3195/2014) via its Science Without Borders program. ALRB acknowledge financial support from Brazilian government agency FACEPE.
 
\bibliography{ref,library}

\end{document}